\documentclass{article}
\usepackage{graphicx} 
\usepackage[strict]{changepage}
\usepackage{framed}
\usepackage{xcolor}
\usepackage{mdframed}
\usepackage{enumitem}
\usepackage{booktabs}
\usepackage{natbib}
\usepackage{amsmath}
\usepackage{cleveref}
\usepackage{rotating}
\usepackage{url}
\usepackage{amsfonts}
\usepackage[margin=1.5in]{geometry}

\definecolor{formalshade}{rgb}{0.97,0.97,0.97}
\newenvironment{formal}{%
  \MakeFramed{\advance\hsize-\width\FrameRestore}%
  \noindent\hspace{-4.55pt}
  \begin{adjustwidth}{}{7pt}%
  \vspace{2pt}\vspace{2pt}%
}
{%
  \vspace{2pt}\end{adjustwidth}\endMakeFramed%
}

\newcommand{\se}[1]{\textcolor{gray}{\footnotesize #1}}

\title{Reindex-Then-Adapt: Improving Large Language Models for Conversational Recommendation}
\author{Zhankui He$^{1,*}$ \and Zhouhang Xie$^1$ \and Harald Steck$^2$ \and Dawen Liang$^2$ \and Rahul Jha$^2$ \and Nathan Kallus$^{2,3}$ \and Julian McAuley$^1$}
\date{%
    $^1$UC San Diego\quad $^2$Netflix\quad $^3$Cornell University\\[2ex]%
    $^*$\url{zhh004@ucsd.edu}
}

\begin{document}

\maketitle

\begin{abstract}
Large language models (LLMs) are revolutionizing conversational recommender systems by adeptly indexing item content, understanding complex conversational contexts, and generating relevant item titles. However, controlling the distribution of recommended items remains a challenge. This leads to suboptimal performance due to the failure to capture rapidly changing data distributions, such as item popularity, on targeted conversational recommendation platforms. In conversational recommendation, 
LLMs recommend items by generating the titles (as multiple tokens) autoregressively, making it difficult to obtain and control the recommendations over all items.
Thus, we propose a \emph{Reindex-Then-Adapt (RTA)} framework, which converts multi-token item titles into single tokens within LLMs, and then adjusts the probability distributions over these single-token item titles accordingly. The RTA framework marries the benefits of both LLMs and traditional recommender systems (RecSys): understanding complex queries as LLMs do; while efficiently controlling the recommended item distributions in conversational recommendations as traditional RecSys do. Our framework demonstrates improved accuracy metrics across three different conversational recommendation datasets and two adaptation settings.
\end{abstract}

\section{Introduction}
\label{sec:intro}

\begin{figure}[tbp]
  \centering
    \includegraphics[width=
0.8\columnwidth]{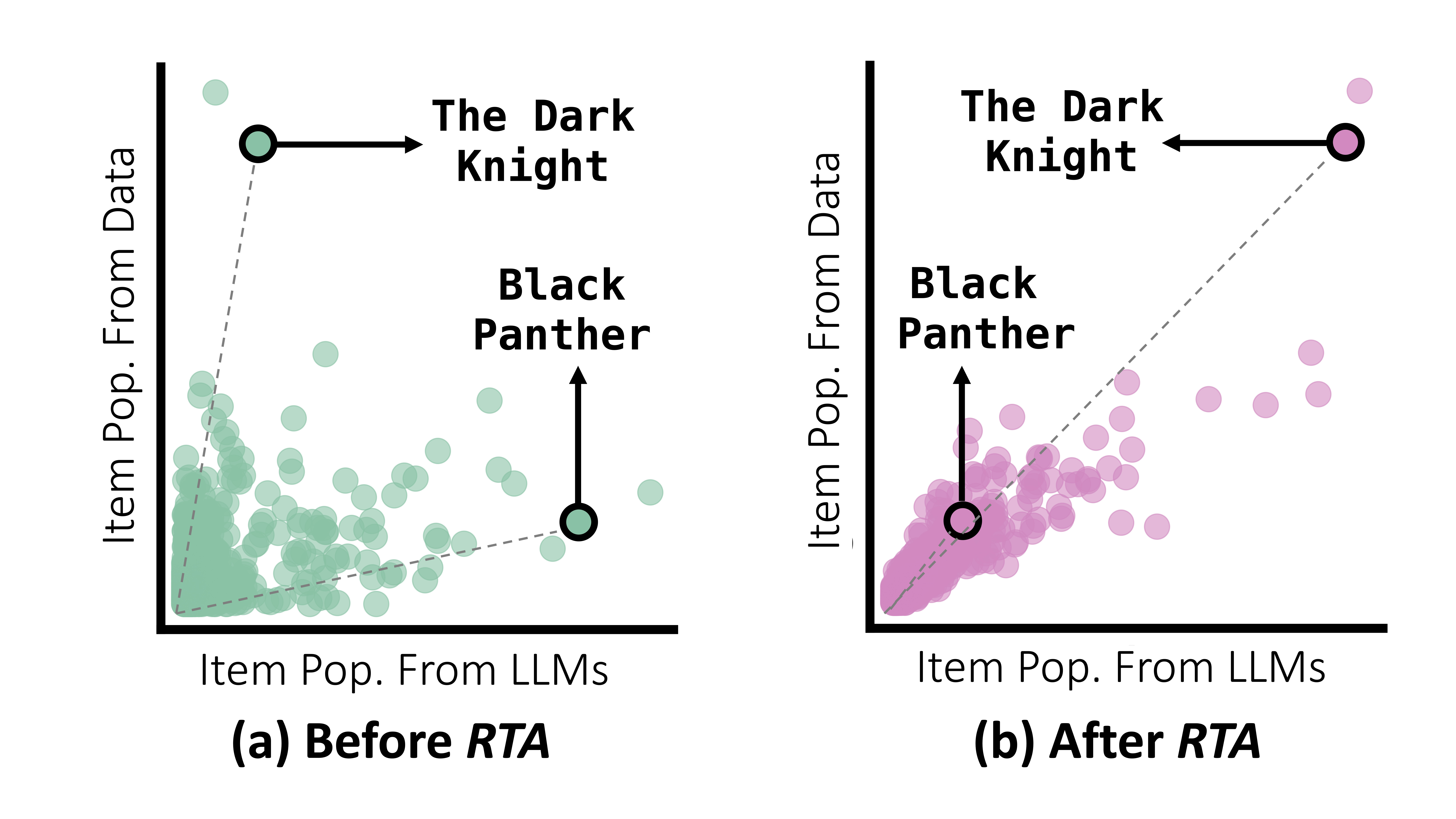}
  \caption{Representative items (\emph{The Dark Knight} and \emph{Black Panther}) demonstrate popularity misalignments between the dataset (ReDIAL~\citep{li2018redial}) and the LLM (Llama2-7b~\citep{touvron2023llama}). This misalignment implies a significant room for recommendation accuracy improvement. Our \emph{Reindex-Then-Adapt (RTA)} framework addresses this gap by aligning the distributions (e.g., item popularites), leading to substantial accuracy improvements.
}
  \label{fig:intro}
\end{figure}

Conversational Recommender Systems~\citep{christakopoulou2016towards,li2018redial,he2023large} (CRS) are an emerging recommendation task aiming to suggest relevant and personalized items via interactive dialogues between users and systems. Recently, Large Language Models~\citep{openai2022chatgpt,brown2020gpt3,he2023large,feng2023large,kang2023llms} (LLMs) have demonstrated proficiency in understanding user intentions within natural language conversational contexts and exhibited substantial domain-specific knowledge (e.g., in movies). Consequently, LLMs offer distinct advantages for CRS and outperform existing non-LLM baselines~\citep{he2023large,wang2023rethinking,feng2023large}. This has garnered significant interest within the research community, positioning LLMs as an indispensable component of CRS.




In this work, we first provide preliminary analysis for LLMs as conversational recommenders. In detail, we view LLMs for conversational recommendations as Differentiable Search Indexing (DSI)~\citep{tay2022transformer, chen2023understanding} models, then study LLMs' \textbf{abilities} and \textbf{limitations} for \textit{item-indexing} and \textit{item-recommendation} tasks:

\begin{itemize}

\item \textbf{Abilities}: LLMs have indexed numerous popular movies, potentially adequate to understand complex conversation contexts and address many movie conversational recommendation scenarios.

\item \textbf{Limitations}: LLMs exhibit misalignment with data distributions from target platforms, as illustrated by item popularity in~\Cref{fig:intro}. Moreover, data distributions like item popularity evolve rapidly in practice, making adjusting LLMs more challenging.

\end{itemize}

We propose to overcome the aforementioned misalignment limitation by easily adjusting LLMs towards changing target distributions. \Cref{fig:intro} illustrates an example with an LLM, Llama-7b~\citep{touvron2023llama}, on the conversational recommendation dataset, ReDIAL~\citep{li2018redial}. Despite the promising conversational recommendations by LLMs~\citep{he2023large}, \Cref{fig:intro}(a) points out a lack of alignment with the data distribution of the target recommendation platform. For example, \emph{The Dark Knight} is popular on ReDIAL~\citep{li2018redial} but not within the LLM, while \emph{Black Panther} presents a contrasting scenario. Our proposed approach alleviates this issue by adjusting the recommendation distributions for all target items from LLMs. \Cref{fig:intro}(b) shows our approach results in more aligned item popularity between LLMs and the target dataset or platform for recommended items such as \emph{The Dark Knight} and \emph{Black Panther}. These alignments bring additional recommendation accuracy improvements, and may have broader benefits, including controllability and fairness.


To achieve this recommendation probability distribution adjustment, there exists a technical challenge. Unlike adjusting recommendation probability distributions over all target items via tweaking the \textit{logit} vectors in traditional RecSys, obtaining such \textit{logit} vectors from LLMs is challenging due to their \textit{generative retrieval} paradigm for CRS~\citep{he2023large}. LLMs generate recommendations by auto-regressively producing multiple item titles (e.g., Top-10), represented by varying numbers of tokens. This process makes obtaining probability distributions over all recommended items computationally expensive, hindering subsequent control or adjustment efforts.

To overcome this challenge, we propose a \emph{Reindex-Then-Adapt (RTA)} framework. With treating LLMs as DSI models for conversational recommendations, we first conduct a \textbf{\texttt{reindex step}}: for original LLMs, we convert already-indexed multi-token item titles (e.g., \emph{Edge of Tomorrow}) into single tokens (e.g., \texttt{|Edge\_of\_Tomorrow|}); then we conduct an \textbf{\texttt{adapt step}}: for reindexed LLMs, the recommended item distributions can be obtained efficiently for following \texttt{adapt} step, e.g., \textit{bias terms adjustment} or \textit{RecSys gating}.

Based on the \emph{RTA} framework, we investigate four reindexing modules and two adaptation strategies across three CRS datasets. Our experimental results show improved recommendation accuracy in CRS. For instance, we improve the recommendation accuracy for the original Llama2-7b by 59.37\% in terms of the Top-10 Hit Rate, surpassing all open-source baselines. Additionally, our studies highlight the significance of adjusting LLMs towards target distributions in CRS and provide insights into scheduling conversational recommendation modules with LLMs.





   \section{Preliminaries}
\label{sec:preliminaries}

\subsection{Task Formulation}

In CRS, a conversation is represented by $C = (u_t, s_t, \mathbf{I}_t)^T_{t=1}$ involving users $u_t \in \mathcal{U}$ and items $\mathbf{I}_t \subseteq \mathcal{I}$ with $T$ conversation turns. Each utterance $s_t$ comprises tokens $v_i$ from vocabulary $\mathcal{V}$. A conversation typically involves a \textit{seeker} and a \textit{recommender}. As formulated in CRS studies ~\citep{li2018redial, chen2019kbrd, zhou2020kgsf, wang2022unicrs,he2023large}, our goal is to learn a recommender to generate a ranked list of items $\hat{\mathcal{I}}_k$ at turn $k$ that aligns with $\mathcal{I}_k$, based on the preceding context $(u_t, s_t, \mathcal{I}_t)^{k-1}_{t=1}$.

\begin{figure}[tbp]
  \centering
    \includegraphics[width=0.9\columnwidth]{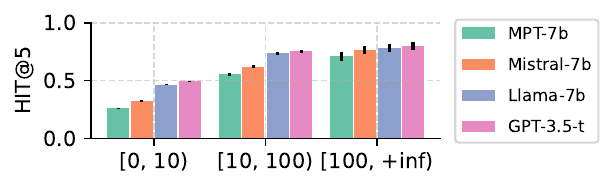}
  \caption{Item Indexing tasks by using movie descriptions from WikiPedia~\citep{auer2007dbpedia} to prompt movie titles. We tested \texttt{MPT-7b}, \texttt{Mistrial-7b} , \texttt{Llama2-7b} and \texttt{GPT-Turbo-3.5} models and group the accuracy by the range of occurrences of the movies in ReDIAL CRS dataset~\citep{li2018redial}. We measure the performance with HIT@5, i.e., whether the target movie in the top-K movie list generated by the LLMs to reflect the movie knowledge stored in the parameters of the LLMs.}
  \label{fig:item-indexing}
\end{figure}

\subsection{Differentiable Search Index (DSI)}

The transformer model has shown proficiency in retrieval tasks, encoding item information within its parameters, which is a method termed Differentiable Search Index (DSI)~\citep{tay2022transformer}. DSI involves two key training tasks for pre-trained language models: \emph{Learn to Index} (\texttt{L2I}) and \emph{Learn to Retrieve} (\texttt{L2R}), 
which can be used to train a model jointly or in a sequential order. \texttt{L2I} focuses on mapping item content, such as movie description, to item indices, exemplified by linking a description of \textit{Edge of Tomorrow} to its title:

\begin{formal}
  \small
  \textbf{L2I Example}: \emph{``A 2014 American science fiction action film starring Tom Cruise and Emily Blunt with ...''} $\rightarrow$ \textbf{Edge of Tomorrow}
\end{formal}

\texttt{L2R}, on the other hand, maps queries to item indices, such as:

\begin{formal}
  \small
  \textbf{L2R Example}: \emph{``I'm feeling bored today and looking for a sci-fi action movie, preferably starring Tom Cruise.''} $\rightarrow$ \textbf{Edge of Tomorrow}
\end{formal}

DSI models are originally proposed for text retrieval tasks~\citep{tay2022transformer}, yet their formulation can be connected to LLMs used in CRS. Considering the \emph{LLMs as CRS} framework proposed in~\citep{he2023large} through the lens of DSI, we observe that:
\begin{itemize}
    \item \textbf{Item Indexing}: LLMs index items by using the item titles (e.g., \emph{``Edge of Tomorrow''}) as the item identifiers via \texttt{L2I}.
    \item \textbf{Item Recommendation}: LLMs use conversational context as queries to generate item indices via \texttt{L2R}.
\end{itemize}

Thus, LLMs inherently function as DSI models, by including a certain number of training samples for \texttt{L2I} and \texttt{L2R} tasks in their pre-training corpus. Compared to common two-tower models, DSI models require only a single model for item recommendations, by indexing item information into its parameters~\citep{tay2022transformer}. 


\subsection{Item Indexing: LLMs Show Sufficient Item Content Knowledge}
\label{sec:indexing}

According to~\citep{he2023large}, LLMs demonstrate superior knowledge in content and context, particularly in the movie domain. This proficiency is attributed to the performance in the ``Learn to Index'' (\texttt{L2I}) task, as viewed through the DSI~\citep{tay2022transformer}. Therefore, our primary concern is the extent to which item content has been indexed in LLMs through the pre-training corpus.

\begin{table}[tbp]
\caption{Additional statistics for cold items, warm items and popular items in ReDIAL~\citep{li2018redial} datasets. \#Items counts the total numbers of items and \#Occurences sums the total occurences of such items in conversations.}
\small
\centering
\begin{tabular}{lccc}
\toprule
   \textbf{ReDIAL}                   & \textbf{Cold -- [0, 10)}       & \textbf{Warm -- [10, 100)}      & \textbf{Pop. -- [100, +inf)}    \\ \midrule
\textbf{\#Items}      & \textbf{4,960 (78.97\%)} & 1,193 (18.99\%)           & 128 (2.04\%)              \\
\textbf{\#Occurences} & 12,523 (18.03\%)         & \textbf{33,304 (47.94\%)} & \textbf{23,647 (34.04\%)} \\ \bottomrule
\end{tabular}
\label{tab:item-freq}
\end{table}

\subsubsection{Observation}

We gathered 6,281 pairs of movie titles from ReDIAL~\citep{li2018redial} and the related descriptions from Wikipedia\footnote{\url{https://www.wikipedia.org/}} for experiments in~\Cref{fig:item-indexing} to assess LLMs performance on \texttt{L2I} task, and observe that:

\begin{itemize}
    \item \textbf{Good Content Knowledge for Popular Items}: All LLMs had indexed a considerable amount of movie content for conversational recommendation tasks. Notably, for frequently mentioned movies,  as defined as movie occurences in the ReDIAL dataset~\citep{li2018redial} that the item occurences range is [100, +inf), all LLMs exhibit impressive content knowledge.
    \item \textbf{Best LLMs}: The proprietary model GPT-3.5-t~\citep{openai2022chatgpt} outperforms others. Among the open-sourced LLMs of similar size, Llama2 demonstrates the best performance in the given task, as shown in \Cref{fig:item-indexing}, making it the chosen base model for our subsequent experiments.
\end{itemize}

\subsubsection{Impact} \Cref{fig:item-indexing} shows that, in terms of item indexing capability as DSI models, LLMs without specific fine-tuning have already indexed numerous popular movies. \Cref{tab:item-freq} shows the imbalance of item occurrences in conversational recommendations, where items labeled as \textit{warm} and \textit{pop.} constitute about 20\% in terms of item counts but contribute to over 80\% of occurrences. This suggests that \textbf{zero-shot LLMs may be sufficient for handling many movie conversational recommendation scenarios}, because many are about warm or popular movies. Although, we admit fine-tuning LLMs to cover more cold items remains future work.

\begin{figure}[tbp]
  \centering
    \includegraphics[width=
0.9\columnwidth]{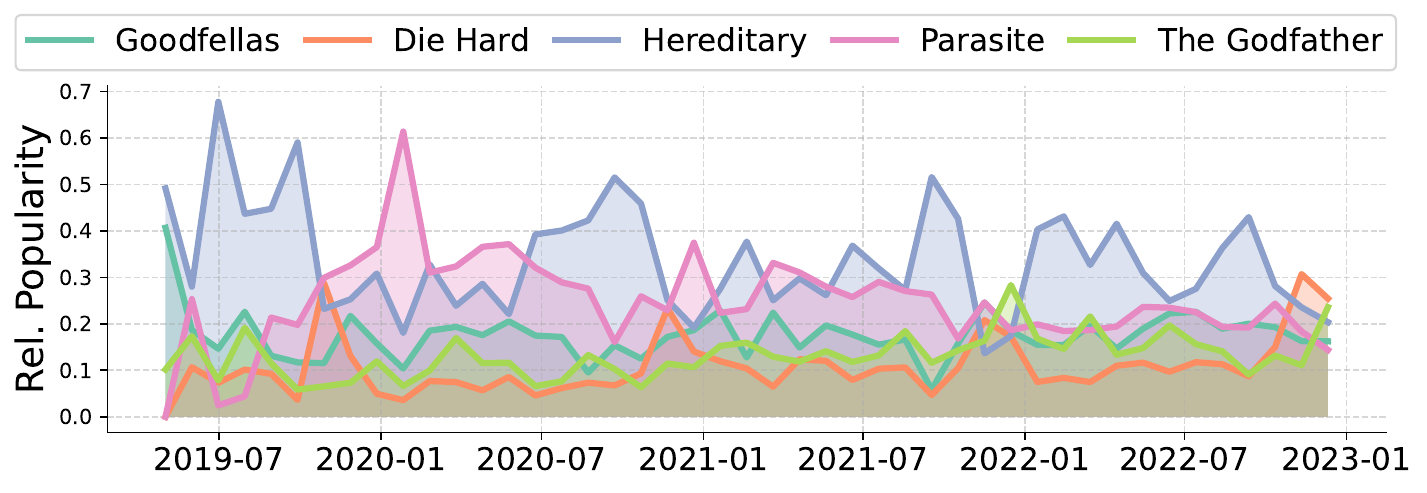}
  \caption{Visualization of item monthly relative popularity from Reddit-Movie~\citep{he2023large} datasets, since this dataset is the only CRS dataset with long-range timestamps in the wild. Item popularities are shown changing overtime rapidly.}
  \label{fig:item-time}
\end{figure}

\subsection{Item Recommendation: LLMs Show Severe Distribution Misalignment}
\label{sec:item_rec}

\subsubsection{Observation}

In this section, we aim to show that even though LLMs can index items effectively, the data distributions LLMs fitted in training do not match the target inference distributions for CRS. We discuss this \emph{distribution misaligment} from two perspectives, using item popularity distribution as an example:

\begin{itemize}
    \item \textbf{Static Perspective:} As depicted in \Cref{fig:intro}(a), LLMs reflect item popularities from the large-scale training corpus to some extent, which often do not align with popular items on the specific platform for CRS.
    \item \textbf{Dynamic Perspective:} Target data distributions, such as item popularity, undergo rapid changes over time due to factors like seasons and promotion strategies. For instance, \Cref{fig:item-time} shows that monthly relative item popularities on the \emph{Reddit-Movie}~\citep{he2023large} dataset change over time, which cannot be captured by a static LLM, even fine-tuned ones.
\end{itemize}

\subsubsection{Impact} Our observations highlight the distribution misalignments between items on the target platform and those recommended by LLMs. This misalignment is considered from both static and dynamic perspectives, suggesting that: (1)~despite LLMs exhibiting impressive performance in conversational recommendation\citep{he2023large}, \textbf{there exists room for improving recommendation accuracy by aligning with the distributions of target platforms}; (2)~due to the dynamic nature of recommendation platforms, target data distributions change rapidly, necessitating the \textbf{more efficient methods to adjust item recommendations from LLMs accordingly}.


\begin{figure*}[tbp]
  \centering
    \includegraphics[width=\textwidth]{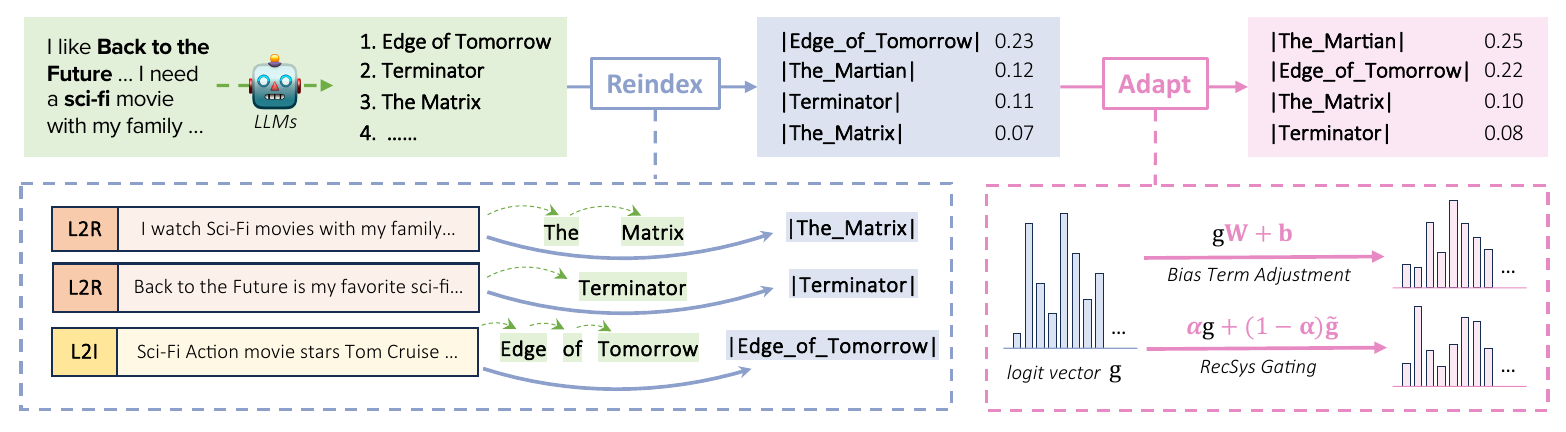}
  \caption{Reindex-Then-Adapt (RTA) Framework. LLMs can generate a list of item titles as the recommendations given the conversation contexts. To further improve the accuracy and controllability, we conduct (1)~\texttt{reindex} step: reindexing the item (e.g., movie) titles in LLMs as single tokens to obtain the predicted \textit{logit} vectors efficiently; (2)~\texttt{adapt} step: adapting the recommenders towards target data distributions effectively with multiple options on the \textit{logit} vectors such as adjusting bias terms or combining RecSys models with Gating mechanism.}
  \label{fig:rta_framework}
\end{figure*}

\section{Framework}

\subsection{Overview}

As discussed in~\Cref{sec:intro}, we argue that representing target items with varying token counts in LLMs poses challenges for adjusting recommendation distributions over all target items. To tackle this, we view LLMs as DSI models that have already indexed sufficient item content knowledge (see~\Cref{sec:indexing}), and propose the \textit{Reindex-Then-Adapt (RTA)} framework, illustrated in~\Cref{fig:rta_framework}:
\begin{enumerate}
    \item \textbf{Reindex} item indices with varying token numbers in LLMs into single-token item indices using a mixture of data samples from the \texttt{L2I} corpus and/or \texttt{L2R} corpus. This aims to remove the \texttt{adapt} step barrier. In contrast to the \texttt{index} step in the original DSI models~\citep{tay2022transformer}, the \texttt{\textbf{re}index} step reuses the content of indexed items from the LLMs, thereby facilitating the learning process for the new item indices.
    
    \item \textbf{Adapt} \textit{logits} from the reindexed LLMs, achieved by transforming the \textit{logit} vectors or by combining with other traditional RecSys using Gating mechanism~\citep{hochreiter1997long,chung2014empirical,gu2020improving}. This adjustment aims to effectively align the recommendation probability distributions over items with the target data distributions.
\end{enumerate}

\subsection{\texttt{Reindex} Step: Single-Token Items in LLMs}

The key of \texttt{reindex} step is to ``squeeze'' multi-token item embeddings into single-token item embeddings efficiently, and preserves the semantics of the original item embeddings in LLMs generation. 

\subsubsection{Identify Item Indices.}
Formally, for a sentence $s=(v_i)^m_{i=1}$ consisting of tokens $v_i \in \mathcal{V}$, we denote the tokens representing an item for CRS tasks as $(v_i)^{j+n}_{i=j}$, where $j$ indicates the starting position of the first token for the item in $s$, and $n$ is the number of tokens representing this item. Consequently, with the \texttt{Embed} layer, LLMs can retrieve the sequence of token embeddings for this item:
\begin{equation}
    \left(\mathbf{v}_i\right)^{j+n}_{i=j} = \texttt{Embed}\left((v_i)^{j+n}_{i=j}\right),
\end{equation}
where $\mathbf{v} \in \mathbb{R}^d$ is the token embedding, and $d$ is the embedding size. 
For example, we may look up \emph{``Edge of Tomorrow''} embeddings represented by ``$[14,72,98]$'' in the sentence $s$.

\subsubsection{Aggregate Multi-Token Embeddings} In the proposed \texttt{reindex} step, we assume that the semantics from multiple (typically shorter than 10) token embeddings can be aggregated into a new single token embedding with a trainable aggregator, such as:
\begin{equation}
    \tilde{\mathbf{v}} = \texttt{Aggregator}\left(\left(\mathbf{v}_i\right)^{j+n}_{i=j}\right),
\end{equation}
where the aggregated $\tilde{\mathbf{v}} \in \mathbb{R}^d$ serves as the new representation of the target item in LLMs generation. Therefore, if all target items are represented by the ``squeezed'' single embedding, scoring all the target items to obtain a logit vector from LLMs is efficient. In general, many existing model architectures can be used as \texttt{Aggregator}, such as RNN-based~\citep{cho2014properties}, Transformer-based models~\citep{vaswani2017attention}, or even Weighted Pooling. We discuss the details and the comparisons with pure new embeddings in~\Cref{sec:aggregator}.
\label{sec:aggregating}

\subsubsection{Learning Process}




The contrative loss~\citep{oord2018representation} is used for learning the aggregator to ``squeeze'' multi-token item embeddings and preserve the semantics for LLMs:
\begin{equation}
    \mathcal{L}_{\textit{reindex}}=-\frac{1}{|\mathcal{D}|}\sum_{\mathbf{q}, \mathbf{\tilde{v}}\in\mathbf{\mathcal{D}}} \log\left[ \frac{\exp \left(\mathbf{q}^\top\mathbf{\tilde{v}}\right)}{\exp \left(\mathbf{q}^\top\mathbf{\tilde{v}}\right)+\sum_{\mathbf{n}\in\mathcal{N}} \exp \left(\mathbf{q}^\top\mathbf{n}\right)}\right],
    \label{eq:reindex}
\end{equation}
where we loop over the training set $\mathcal{D}$ that consists of $(\mathbf{q}, \tilde{\mathbf{v}})$ pairs. Those pairs are collected from sentences containing target items. Here, $\mathbf{q}\in\mathbb{R}^d$ is the contextual embedding of the last position from a LLM, which is originally used to generate the first token of original indexed items, but now we aim to force the aggregated item representation $\mathbf{\tilde{v}}\in\mathbb{R}^d$ described in~\Cref{sec:aggregating} to be generated by LLMs. To achieve this reindex step, we also prepare negatives $\mathbf{n}\in\mathbb{R}^d$ from the negative representation set $\mathcal{N}$. 

Two groups of corpus are considered in the \texttt{reindex} step:

\begin{itemize}
    \item \textbf{L2R Data}: In this training corpus, (query, target) pairs  are used as \texttt{L2R} samples. In CRS context, those are samples from the conversations.
    \item \textbf{L2I Data}: In this training corpus, (content, target) pairs are used as \texttt{L2I} samples. In CRS context, those are samples from the item metadata like textual descriptions.
    \item \textbf{Data Mixture}: In this case, we consider mixing the both \texttt{L2R} and \texttt{L2I} samples as a unified corpus and use it to train our model jointly. We use this option and include details in~\Cref{app:implementation}.
\end{itemize}



\subsection{\texttt{Adapt} Step: Item Probabilities Adjustment}

After re-indexing, all the items are represented by single-token embeddings. It makes recommendation as easy as one-step decoding in LLMs, and also enables multiple efficient ways to adjust the recommendation item distributions to adapt towards target platforms or specific data distributions. We introduce two types of adaptation methods in the following sections, one for item popularity adjustments, and another one for combining with the traditional recommender systems. To start with, we assume the logit vector $\mathbf{g} \in \mathbb{R}^{|\mathcal{I}|}$ has already been given by the LLMs, and the corresponding probability vector should be $\mathbf{p} = \texttt{softmax}(\mathbf{g})$.

\subsubsection{Bias Term Adjustment} Inspired by~\citep{zhao2021calibrate}, a common way to adjust logits is an affine transformation, i.e.:
\begin{equation}
    \hat{\mathbf{p}} = \texttt{softmax}\left(\mathbf{g}\mathbf{W} + \mathbf{b}\right),
\end{equation}
where $\mathbf{W}\in\mathbb{R}^{|\mathcal{I}|\times |\mathcal{I}|}$ is a weight matrix and $\mathbf{b}\in\mathbb{R}^{|\mathcal{I}|}$ is the bias term. Similar to~\citep{zhao2021calibrate}, we restrict the matrix $\mathbf{W}$ to be diagonal to prevent the size of parameters from growing quadratically in the size of items. Therefore, in this special case, we are able to interpret the $\mathbf{W}$ and $\mathbf{b}$ as \textit{multiplicative} and \textit{additive bias} terms towards the target data distributions, respectively.

\subsubsection{Traditional RecSys Gating} Inspired by~\citep{he2023large}, we notice that LLMs excel at content/context knowledge, but traditional RecSys, where the output logit vector can be denoted as $\tilde{\mathbf{g}} \in \mathbb{R}^{|\mathcal{I}|}$, is good at collaborative knowledge instead. Motivated by this observation, combining those two types of models becomes easy after the re-indexing step:
\begin{equation}
    \hat{\mathbf{p}} = \texttt{softmax}\left(\alpha \mathbf{g} + (1-\alpha)\tilde{\mathbf{g}}\right),
\end{equation}
where the coefficient $\alpha \in [0, 1]$ can be set in many different ways. For simplicity sake, we use a learnable scalar $\tilde{\alpha}$ for $\alpha = \texttt{sigmoid}\left(\tilde{\alpha}\right)$ in our experiments, but more options can be considered, such as being predicted by a MLP model to naturally determine how much we should weight the responses from LLMs or traditional RecSys, like $\alpha = \texttt{sigmoid}\left(\texttt{MLP}(\mathbf{q})\right)$, where $\mathbf{q}\in \mathbb{R}^d$ can be the contextual embedding from a LLM in~\Cref{eq:reindex}.

\subsubsection{Learning Process} 

We use maximum likelihood estimation to derive the loss for \texttt{adapt} step, in order to learn the parameters of the bias terms or the recsys model. Note that the LLMs parameters are not involved in this step, ensuring an efficient learning process:
\begin{equation}
\mathcal{L}_{\textit{adapt}}=-\frac{1}{|\mathcal{D}^*|}\sum_{i=1}^{|\mathcal{D}^*|} \log \hat{\mathbf{p}}_{i,*},
\end{equation}
where dataset $\mathcal{D}^*$ is collected from the target platform, such as ReDIAL~\citep{li2018redial}, which is typically a small sized dataset. Here, $\hat{\mathbf{p}}_{i,*}$ denotes the probability of the ground-truth item in the $i^\text{th}$ data sample. Our purpose is to adapt the model towards the underlying data distributions of $\mathcal{D}^*$ through this learning process.

\begin{table}[t]
\caption{Dataset Statistics. We update \emph{Reddit-Movie} CRS dataset as Reddit-V1.5 according to the raw data dump provided by~\citep{he2023large} from 2012 to 2022. Specifically, conversation turns with valid recommended items are denoted as \emph{R\_Turns}. }
\centering
\resizebox{0.8\columnwidth}{!}{
\begin{tabular}{lccc}
\toprule
                    & \textbf{INSPIRED} & \textbf{ReDIAL} & \textbf{Reddit-V1.5} \\ 
                    & \citep{hayati2020inspired} & \citep{li2018redial} & \citep{he2023large} \\ 
                    \midrule
\textbf{\#Conv.}    & 999               & 11,348          & 2,726,471                  \\
\textbf{\#Turns}    & 21,124            & 139,557         & 5,063,007                  \\
\textbf{\#R\_Turns} & 1,950             & 30,322          & 1,787,050                  \\
\textbf{\#Users}    & 999               & 764             & 520,913                    \\
\textbf{\#Items}    & 1,472             & 6,281           & 68,285                     \\ \bottomrule
\end{tabular}
}
\label{tab:data}
\end{table}

\begin{table}[]
\centering
\caption{The main results for our models on conversational recommendation accuracy performance, compared against (1) traditional recommendation models; (2) zero-shot large language models (LLMs); (3) traditional conversational recommendation models; and (4) zero-shot dense retrievers. The size of the reported LLMs used here is 7B. We denote the model metrics with the best performance in bold. Llama2-R denotes the Llama2-7b model after our \texttt{reindex} step. We also show the results after the \texttt{adapt} step with bias terms (+Bias) or RecSys model combination with Gating mechanism (+RecSys). }
\resizebox{\textwidth}{!}{
\begin{tabular}{@{}lcccccccccccc@{}}
\toprule
                    & \multicolumn{4}{c}{\textbf{INSPIRED}}                         & \multicolumn{4}{c}{\textbf{ReDIAL}}                           & \multicolumn{4}{c}{\textbf{Reddit-V1.5}}                           \\ \cmidrule(l){2-5} \cmidrule(l){6-9} \cmidrule(l){10-13}
    & H@5           & N@5           & H@10          & N@10          & H@5           & N@5           & H@10          & N@10          & H@5           & N@5           & H@10          & N@10          \\ \midrule
\textbf{Popuplarity}        & {.089 \se{.020}}           & {.065 \se{.015}}           & .103 \se{.021}            & {.070 \se{.015}}            & .035 \se{.003}           & .025 \se{.002}           & .052 \se{.003}            & .030 \se{.002}            & .008 \se{.001}           & .004 \se{.000}           & .014 \se{.001}            & .006 \se{.000}            \\
\textbf{FISM}      & .075 \se{.018}           & .045 \se{.012}           & .103 \se{.021}            & .054 \se{.012}            & .065 \se{.004}           & .040 \se{.003}           & .112 \se{.005}            & .054 \se{.003}            & .022 \se{.001}           & .012 \se{.001}           & .043 \se{.001}            & .019 \se{.001}            \\ 
\textbf{SASRec}       & .061 \se{.016}           & .037 \se{.010}           & .103 \se{.021}            & .051 \se{.011}            & .068 \se{.004}           & .041 \se{.002}           & .116 \se{.005}            & .056 \se{.003}     &    .022 \se{.001}           & .013 \se{.001}           & .039 \se{.001}            & .018 \se{.001}            \\  \hline
\textbf{MPT}     & .075 \se{.018}           & .045 \se{.011}           & .099 \se{.020}            & .052 \se{.012}            & .072 \se{.004}           & .045 \se{.003}           & .116 \se{.005}            & .059 \se{.003}            & .026 \se{.001}           & .017 \se{.001}           & .040 \se{.001}            & .021 \se{.001}            \\
\textbf{Mistral} & .061 \se{.016}           & .040 \se{.011}           & .066 \se{.017}            & .041 \se{.012}            & .082 \se{.004}           & {.056 \se{.003}}           & .111 \se{.005}            & .065 \se{.003}            & .029 \se{.001}           & .020 \se{.001}           & .038 \se{.001}            & .023 \se{.001}            \\
\textbf{Llama2}  & .080 \se{.019}           & .050 \se{.012}           & .122 \se{.022}            & .064 \se{.013}            & \textbf{.094 \se{.004}}           & {.059 \se{.003}}           & {.145 \se{.005}}            & {.075 \se{.003}}            & .042 \se{.001}           & .027 \se{.001}           & .064 \se{.001}            & .034 \se{.001}            \\ \hline
\textbf{ReDIAL}     & .060 \se{.016}           & .041 \se{.012}           & .106 \se{.021}            & .056 \se{.012}            & .067 \se{.004}           & .044 \se{.003}           & .106 \se{.005}            & .057 \se{.003}            & .029 \se{.001}           & .019 \se{.001}           & .044 \se{.001}            & .024 \se{.001}            \\
\textbf{UniCRS}    &      {.091 \se{.019}}                             &     {.055 \se{.011}}                              &      {.132 \se{.019}}                             &              {.073 \se{.014}}                     &      {.085 \se{.003}}                             &           {.058 \se{.003}}                        &           {.112 \se{.004}}                        &          {.071 \se{.003}}                         &         {.028 \se{.001}}                          &              {.017 \se{.001}}                     &             {.040 \se{.001}}                      &         {.021 \se{.001}}                          \\ \midrule
\textbf{SBERT}    & .038 \se{.013}           & .026 \se{.010}           & .066 \se{.017}            & .036 \se{.010}            & .016 \se{.002}           & .010 \se{.001}           & .026 \se{.002}            & .013 \se{.001}            & .003 \se{.000}           & .002 \se{.000}           & .005 \se{.000}            & .002 \se{.000}            \\
\textbf{Instructor}    & .052 \se{.015}           & .034 \se{.011}           & .085 \se{.019}            & .045 \se{.011}            & .025 \se{.002}           & .013 \se{.001}           & .043 \se{.003}            & .019 \se{.001}            & .009 \se{.001}           & .006 \se{.000}           & .017 \se{.001}            & .008 \se{.000}            \\ \midrule
\textbf{Llama2-R}    & .066 \se{.017}           & .041 \se{.011}           & .103 \se{.021}            & .053 \se{.012}            & .071 \se{.004}           & .042 \se{.002}           & .117 \se{.005}            & .057 \se{.003}            & .055 \se{.001}           & .035 \se{.001}\textbf{}           & .093 \se{.002}            & .047 \se{.001}            \\
\textbf{  +Bias}     & \textbf{.103 \se{.021}}           & \textbf{.066 \se{.014}}           & \textbf{.164 \se{.025}}            & \textbf{.083 \se{.015}}            & .083 \se{.004}           & .053 \se{.003}           & .123 \se{.005}            & .066 \se{.003}            & {.059 \se{.001}}           & {.037 \se{.001}}           & {.096 \se{.002}}            & {.049 \se{.001}}                                  \\ 
\textbf{  +RecSys}      & .089 \se{.020}           & .052 \se{.013}           & \textbf{.164 \se{.025}}            & .076 \se{.013}            & \textbf{.094 \se{.004}}           & \textbf{.060} \se{.003}           & \textbf{.146 \se{.005}}            & \textbf{.076 \se{.003}}            &    \textbf{.061 \se{.001}}                     &       \textbf{.038 \se{.001}}                  &     \textbf{.101 \se{.002} }                    &     \textbf{.051 \se{.001}}                    \\
\bottomrule
\end{tabular}
}
\label{tab:main}
\end{table}

\section{Experiments}

\subsection{Experiment Setup}

\subsubsection{Datasets}

Three conversational recommendation datasets~\citep{hayati2020inspired, li2018redial, he2023large} are used in our experiments, where the statistics are summarized in~\Cref{tab:data}: \textbf{INSPIRED}~\citep{hayati2020inspired} and \textbf{ReDIAL}~\citep{li2018redial}: These two datasets consist of small-scale human-human conversations for movie recommendations with crowd-sourced annotations from MTurk~\footnote{\url{https://mturk.com}}. Due to their short collection time span, temporal patterns are unlikely to be observed. Nevertheless, considering their widespread use, we present our model results based on these datasets. In the following experiments, we randomly split the datasets into training, validation, and test sets using an 8:1:1 ratio. \textbf{Reddit-V1.5}~\citep{he2023large}: This dataset comprises large-scale movie discussions on Reddit, which were collected and processed by~\citep{he2023large}. This dataset shows real movie conversation recommendations in the wild and includes corresponding timestamps for 10 years to study temporal patterns. For data splitting, we use the last two months (i.e., Nov. and Dec. in 2022) as validation and testing set respectively to approximate the real setting. Due to the large size of the given dataset, we uniformly sample 20\% conversation turns for validation (i.e., 11,241 samples) and testing (i.e., 13,816 samples). 

\subsubsection{Baselines}

We consider four groups of baseline models for comparison. (1)~We consider some representative traditional \textit{item-based}\footnote{We only use item-based models since INSPIRED does not have historical user interactions.} RecSys models, including \textbf{Popularity}, \textbf{FISM}~\citep{kabbur2013fism} and \textbf{SASRec}~\citep{kang2018self}. (2)~We consider some representative CRS models: \textbf{ReDIAL}~\citep{li2018redial} and \textbf{UniCRS}~\citep{wang2022unicrs}: This model uses a pre-trained language model. (3)~We consider some dense retrieval models given the connections to document retrieval: \textbf{SBERT}~\citep{reimers2019sentence} and \textbf{Instructor}~\citep{su2022one}. (4)~We consider some zero-shot open-sourced LLMs as baselines like~\citep{he2023large} and use the 7-billion-parameter version due to compute burden: \textbf{MPT-7b}~\citep{MosaicML2023MPT}, \textbf{Mistral-7b}~\citep{jiang2023mistral} and \textbf{Llama2-7b}~\citep{touvron2023llama}. We also discuss the results from \textbf{GPT-3.5-turbo}~\citep{openai2022chatgpt}, which is a much larger proprietary model that can achieve state-of-the-art CRS performance even in a zero-shot setting~\citep{he2023large}. The details of baseline models are found in~\Cref{app:baseline}. 

\subsubsection{Evaluation Metrics}

We focus on recommendation accuracy using HIT@K (H@K) and NDCG@K (N@K),  following~\citep{li2018redial, chen2019kbrd, zhou2020kgsf, wang2022unicrs}. We consider the means and the standard errors\footnote{We use error bars in our figures and gray numbers in our tables for standard errors.} of the metrics with $K=\{5,10\}$. Please find the implementation details in~\Cref{app:implementation}.

\subsection{General CRS Performance}

\subsubsection{Baseline Performance.} \Cref{tab:main} shows the recommendation accuracy of four groups of baselines on three conversational recommendation datasets. There are some observations:

\textbf{On Traditional RecSys.} Conventional recsys models effectively capture target popularity and further item-item similarities, resulting in reasonable recommendation accuracies. Interestingly, on INSPIRED, we find that non-personalized popularity serves as a strong baseline, because the limited size of the training set may restrict the ability to capture more complex item-item relationships. The results of traditional recommendation system models also indicate the potential of improving the recommendation accuracy by aligning with target data distributions.

\textbf{On LLMs.} LLMs with zero-shot prompting from~\citep{he2023large} achieve impressive results, surpassing even the best results on ReDIAL datasets. Additionally, the rank of recommendation accuracy within the LLM group aligns with the performance from~\Cref{fig:item-indexing}. Further details on the specific proprietary model \texttt{GPT-3.5-t} are discussed in~\Cref{sec:further}.

\textbf{On Other Baselines.} We observe that zero-shot state-of-the-art dense retrievers are unable to achieve comparable performance as zero-shot LLMs; this may be due to two reasons:  (1)~Dense retrievers focus more on retrieving similar documents according to semantic similarities (e.g.,~similar contents), but LLMs show better understanding abilities for conversation contexts; (2)~We are encoding the movie textual title rather than the description of the movie for fair comparison, which may limit the dense retrievers' performance. As for traditional CRS models, since we  follow the setting in~\citep{he2023large} to remove ``repeated'' items, many popular CRS models perform relatively weaker in the corrected evaluation protocol.


\subsubsection{Ours vs.~Baselines.} We construct a small-sized aggregator on top of 
Llama2 as an example, then use this aggregator to reindex multi-token movie titles into single-token movie titles as recommendation candidates, namely Llama2-R.

\textbf{On Recommendation Accuracy.} \Cref{tab:main} shows that, following the \texttt{reindex} and \texttt{adapt} steps, our model excels over baselines on INSPIRED and Reddit-V1.5 datasets, achieving the competitive best results on the ReDIAL dataset. Examining the \texttt{reindex} step (Llama2-R) and \texttt{adapt} step (+Bias or +RecSys), we observe a potential performance decrease in the \texttt{reindex} step due to the semantic gap from original token embeddings to the new single token embeddings from the relatively small aggregator. However, our models compensate by capturing the target data distribution through bias terms or traditional RecSys models. A more in-depth analysis of these \texttt{adapt} methods will be discussed in~\Cref{sec:adapt}.

\textbf{On Efficiency and Flexibility.} It is crucial to mention that the aggregator-based methods are around $10\times$ smaller than the corresponding out-of-vocabulary item embedding tables and approximately $233\times$ smaller than the Llama2-7b base model, emphasizing its space efficiency. Additionally, as all movie titles with varying numbers of tokens are "squeezed" into single tokens, our model can rank all items with a single decoding step, making it around $100\times$ faster than the generative retrieval from LLMs to recommend the top-20 items. Moreover, single tokens facilitate easy acquisition of the recommendation item distribution, enhancing flexibility in control or further adjustment of the recommendations.

\begin{figure}[tbp]
  \centering
    \includegraphics[width=0.9\columnwidth]{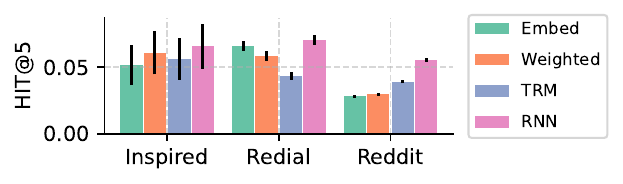}
  \caption{Different methods to represent items in LLMs with single-token embeddings and the related recommendation accuracy HIT@5 after the  \texttt{reindex} step.}
  \label{fig:aggregators}
\end{figure}

\subsection{Effectiveness of the \texttt{Reindex} Step}
\label{sec:aggregator}

\subsubsection{Experiment Setup}
We explore methods for representing item titles with single-token embeddings in LLMs, investigating four approaches: (1)~\textbf{Embed}: randomly initialized out-of-vocabulary (OOV) embeddings. Subsequently, three models aggregate existing LLM token embeddings into a single-token embedding and trained on the samples from those three datasets: (2)~\textbf{Weighted}: learning position-wise attention weights to aggregate multi-token embeddings into a single one, followed by a simple linear projection; (3)~\textbf{TRM}: employing a single-layer transformer to derive a contextual embedding from the output \texttt{CLS} token; (4)~\textbf{RNN}: using a simple GRU model to aggregate multiple token embeddings, with the last hidden state vectors serving as the item representations.

\subsubsection{Embedding vs.~Aggregator} The embedding-based method cannot be shared across different datasets due to the practical challenge in normalizing item titles. However, the aggregators are shared across different datasets, using the raw text of item titles as inputs. \Cref{fig:aggregators} demonstrates that aggregators are not only generalizable across different datasets but also yield superior recommendation accuracy. Interestingly, despite Reddit having a dominant share of training samples (96\%) as shown in~\Cref{tab:data}, the trained aggregators with mixed data samples perform even better than the dataset-specific new embeddings in~\Cref{fig:aggregators}.

\subsubsection{Different Aggregators} Among the three aggregators, the Weighted method demonstrates competitive performance despite its simple architecture. This suggests that the existing token embeddings from the LLMs are effective enough, making the weighted-sum with linear projection a reasonable approach to consolidating token embeddings. Additionally, TRM performs worse than RNN, possibly because (1) titles (e..g,~movies) are typically short (fewer than 20 tokens), diminishing the significance of TRM's advantages over RNN in handling long dependencies; (2) \texttt{CLS} tokens show difficulty in representing a sentence, as noted in the literature~\citep{choi2021evaluation}.

\begin{table}[t]
\caption{Recommendation accuracy comparison among Continual-Training on Llama2-R (Cont.), and the detailed configurations of adding bias terms or RecSys gating.}
\centering
\resizebox{0.9\columnwidth}{!}{
\begin{tabular}{lcccccc}
\toprule
                  & \multicolumn{2}{c}{\textbf{INSPIRED}}               & \multicolumn{2}{c}{\textbf{ReDIAL}}                 & \multicolumn{2}{c}{\textbf{Reddit-V1.5}}            \\ \cmidrule(l){2-3} \cmidrule(l){4-5} \cmidrule(l){6-7} 
                  & H@10                     & N@10                     & H@10                     & N@10                     & H@10                     & N@10                     \\ \midrule
\textbf{Llama2-R} & .103 \se{.021}          & .053 \se{.012}          & .117 \se{.005}          & .057 \se{.003}          & .093 \se{.002}          & .047 \se{.001}          \\
\textbf{Cont.}    & .146 \se{.024}          & .081 \se{.015}          & .124 \se{.004}          & .067 \se{.003}          & .093 \se{.001}          & .047 \se{.001}          \\ \midrule
                  & \multicolumn{6}{c}{\textit{Bias Term Adjustment (+Bias)}}                                                                                                               \\ \cmidrule(l){2-7} 
\textbf{w/ $\mathbf{gW}$}    & .155 \se{.025}          & .081 \se{.014}          & .123 \se{.005}          & .066 \se{.003}          & .093 \se{.001}          & .048 \se{.001}          \\
\textbf{w/ $\mathbf{b}$}     & .103 \se{.021}          & .053 \se{.012}          & .118 \se{.005}          & .057 \se{.003}          & .096 \se{.002}          & .049 \se{.001}          \\
\textbf{w/ $\mathbf{gW+b}$}  & \textbf{.164 \se{.025}} & \textbf{.083 \se{.004}} & .123 \se{.005}          & .066 \se{.003}          & .096 \se{.001}          & .049 \se{.001}          \\ \midrule
                  & \multicolumn{6}{c}{\textit{RecSys Model Gating (+RecSys)}}                                                                                                                \\ \cmidrule(l){2-7} 
\textbf{+ FISM}   & \textbf{.164 \se{.025}} & .076 \se{.013}          & .139 \se{.005}          & .072 \se{.003}          & \textbf{.101 \se{.002}} & .049 \se{.001}          \\
\textbf{+ SASRec} & .136 \se{.023}          & .071 \se{.014}          & \textbf{.146 \se{.005}} & \textbf{.076 \se{.003}} & \textbf{.101 \se{.002}} & \textbf{.051 \se{.001}} \\ \bottomrule
\end{tabular}
}
\label{tab:ablation}
\end{table}

\subsection{Effectiveness of the \texttt{Adapt} Step}
\label{sec:adapt}

\subsubsection{Component Analysis.} \Cref{tab:ablation} shows introducing bias terms after the \texttt{reindex} step is a simple yet effective strategy. This is attributed to the potential for improving recommendation accuracy by addressing popularity misalignments, as discussed in~\Cref{fig:intro}. Additionally, we observe that on the small dataset, INSPIRED, +Bias outperforms +RecSys. This is because the parameter space for learning is significantly reduced, changing from learning item-item relationships to learning item point-wise popularity, which can be effectively captured with a small number of training samples. 

Meanwhile, \Cref{tab:ablation} demonstrates that introducing traditional RecSys models is effective when there is a large number of training samples available to adapt the recommendation distribution. On ReDIAL and Reddit-V1.5, this leads to improved recommendation accuracy compared to Cont.~and +Bias.~However, on the small dataset INSPIRED, using RecSys to learn item-item relationships tends to result in overfitting. This motivates us to consider different \texttt{adapt} steps by cases. For example, after collecting the most recent samples, bias-term adjustment (+Bias) is recommended if the number of new samples is limited. Otherwise, RecSys gating would be a good option.

\subsubsection{Impact of Bias Term Types} Both multiplicative and additive bias terms improve accuracy across diverse datasets, though their impact varies. Specifically, multiplicative bias terms exhibit significant improvement on INSPIRED and ReDIAL datasets, whereas additive bias terms play a pivotal role on Reddit-V1.5. 

\subsubsection{Impact of RecSys Model Types} Our current focus is on "item-based" RecSys models without incorporating long-term user representations. In this context, FISM and SASRec exhibit enhanced performance. Notably, FISM outperforms SASRec on the INPSIRED dataset, possibly due to the complexity of SASRec, a transformer-based model, being less suitable for smaller datasets. Conversely, on larger datasets such as ReDIAL and Reddit-V1.5, SASRec demonstrates superior performance, suggesting that employing transformer-based RecSys models is advantageous when dealing with larger data sizes. Specifically, on ReDIAL, characterized by longer conversation rounds, SASRec may bring additional benefits in capturing item-to-item sequential patterns within conversations.


\begin{figure}[tbp]
  \centering
    \includegraphics[width=0.9\columnwidth]{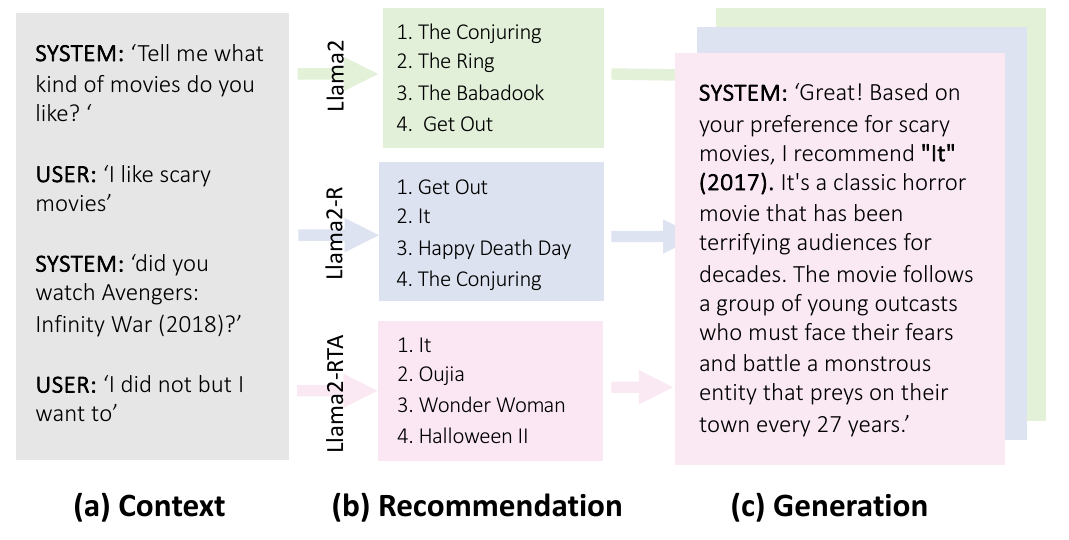}
  \caption{An Example with real results from Llama2, Llama-R and Llama-RTA (+SASRec), followed by a natural language response from Llama2 (detailed prompts can be found in~\Cref{app:prompts}). This conversational context is from ReDIAL dataset, with the ground-truth movie \emph{IT}.}
  \label{fig:example}
\end{figure}

\subsection{Discussions}

\subsubsection{Conversational Recommendation Responses} \Cref{fig:example} illustrates the complete pipeline of generating results for conversational recommendation tasks. Our discussions are below:

\textbf{On Recommendation.} The outputs of the \textit{recommendation} phase are items. In~\Cref{fig:example}, the three models (Llama2 and its variants under our framework) understand contexts, yielding high-quality recommendations for scary movies. Specifically, Llama2-RTA builds a connection between the superhero movie \emph{Avengers: Infinity War} in the context and the candidate \emph{Wonder Woman}, using item-to-item relationships modeled by the SASRec~\citep{kang2018self} model. Meanwhile, we posit that while multiple recommended items align with conversation contexts, the failure to adjust for the popularity of items on the target platform (e.g., movie \emph{IT} being popular on ReDIAL) leads to zero-shot LLMs failing to meet user interests.

\textbf{On Generation.} The outputs of the \textit{generation} phase are texts. In~\Cref{fig:example}, the generation phase is accomplished by prompting the Llama2 model. It is noted that our focus in this work is solely on the technical aspects of the recommendation phase. We treat the generation phase as a separate task that can be completed either by existing LLMs or adjusted based on user interface requirements. Still, we make some observations: (1)~In many cases, presenting the recommendation phase suffices for users. However, our RTA framework, which introduces only a few additional parameters without changing the weights of the original LLMs, efficiently enables the reuse of LLMs for further generating natural-language responses as shown in~\Cref{fig:example}; (2)~In conversational recommendations, there is an ongoing debate about whether to perform the recommendation or generation phase first~\citep{li2018redial,zhou2020kgsf,wang2022unicrs}. Our example suggests that, if the recommendation phase is frequently adjusted (a common scenario due to distribution shift), it is advisable to perform the recommendation phase first and then the generation phase. Reversing the order may lead to text-item inconsistency issues (e.g., the generated response is specifically tailored for recommended movie \emph{IT}, leading to a mismatch with the recommendation from Llama2).


\begin{table}[t]
\caption{Recommendation accuracy comparison our model based on a 7B open-sourced LLM (Llama2) and the proprietary model ChatGPT (GPT-3.5-t).}
\centering
\resizebox{0.8\columnwidth}{!}{
\begin{tabular}{@{}lcccccc@{}}
\toprule
                     & \multicolumn{2}{c}{\textbf{INSPIRED}}                                             & \multicolumn{2}{c}{\textbf{ReDIAL}}                                               & \multicolumn{2}{c}{\textbf{Reddit-V1.5}}                                            \\ \cmidrule(l){2-3} \cmidrule(l){4-5} \cmidrule(l){6-7} 
\textbf{Model} & H@10                                    & N@10                                    & H@10                                    & N@10                                    & H@10                                     & N@10                                     \\ \midrule
\textbf{Ours.}       & .164 \se{.025} & .083 \se{.004} & .131 \se{.005} & .068 \se{.003} &  .102 \se{.002}      &     .052 \se{.001} \\
\textbf{GPT-3.5-t}  &   .150 \se{.024}            & .089 \se{.016}  & .163 \se{.006}            & .089 \se{.003}   & .104 \se{.002}            & .055 \se{.001}    \\ \bottomrule
\end{tabular}
}
\label{tab:gpt}
\end{table}

\subsubsection{Comparison with Proprietary Models} To deepen our understanding of the models, we adopt the setting in~\citep{he2023large} to query the proprietary model {GPT-3.5-t}~\citep{openai2022chatgpt}. As shown in \Cref{tab:gpt}, {GPT-3.5-t} remains a competitive model for conversational recommendations with zero-shot prompting. However, it is reasonable to guess that, given our LLM-architecture-agnostic approach, improving recommendation accuracy based on {GPT-3.5-t} is possible if the weights are accessible. A reasonable next step involves working on models similar to GPT-3.5-t, such as Llama2-70b. This could be pursued as future work, if the required compute resources are available.

\label{sec:further}

   \section{Related Work}

\subsection{Conversational Recommendation (CRS)}
The objective of conversational recommender systems (CRS) is to elicit user preferences and deliver tailored recommendations through interactive dialogues. Historically, CRS implementations have ranged from some template-driven systems~\citep{christakopoulou2016towards,lei2020interactive,lei2020estimation, he2022bundle,zhang2022multiple} to critique-based approaches~\citep{chen2012critiquing,wu2019deep,li2021self}. With the evolution of natural language processing, "deep" CRS models~\citep{li2018redial,chen2019kbrd,wang2022unicrs} have been developed, enabling more natural-language interactions.
Research indicates the utility of CRS models is enhanced by incorporating diverse supplementary data, such as knowledge-enriched models~\citep{chen2019kbrd, zhou2020kgsf} utilizing external knowledge bases~\citep{auer2007dbpedia, liu2004conceptnet}, review-centric models~\citep{lu2021revcore}, and session/sequence-oriented models~\citep{zou2022improving, li2022uccr}.
UniCRS~\citep{wang2022unicrs} uses knowledge bases~\citep{auer2007dbpedia}, built on DialoGPT~\citep{zhang2020dialogpt} and employing prompt tuning~\citep{brown2020gpt3}, represents a state-of-the-art CRS model on datasets like ReDIAL~\citep{li2018redial} and INSPIRED~\citep{hayati2020inspired}. Recently, an emerging topic is to leverage LLMs in CRS, with \citep{friedman2023leveraging,he2023large} introducing a novel CRS pipeline, even in the zero-shot setting~\citep{he2023large}, and \citep{wang2023rethinking} focusing on advanced user simulation for LLM evaluation. Our research is the first to study the distribution misalignments in zero-shot LLMs for CRS and solutions for this issue to improve recommendation accuracy.

\subsection{Large Language Models (LLMs)}

Recent breakthroughs in natural language processing (NLP) have demonstrated that large language models (LLMs) possess a remarkable capacity for generalizing to unfamiliar tasks and areas~\citep{Chowdhery2022PaLMSL, Brown2020LMAreFewShot, Wei2022CotReasoning} in zero-shot or few-shot settings. Studies have shown that scaling up LLMs can significantly enhance their performance and efficiency in downstream applications~\citep{Kaplan2020ScalingLF}. 
In line with these developments, LLMs have been successfully applied to various downstream tasks such as question answering, numerical reasoning, code generation, and commonsense reasoning, often without requiring gradient updates~\citep{zheng2023codegeex,Brown2020LMAreFewShot,Li2022CompetitionlevelCG, Kaplan2020ScalingLF}. The recommendation field has recently begun integrating LLMs, either by adapting LLM architectures~\citep{Geng0FGZ22RecAsLp, cui2022m6rec} or repurposing existing LLMs for recommendation purposes~\citep{li2023gpt4rec, wang2023generative, liu2023chatgpt}. Our study aligns with the research line of utilizing LLMs for conversational recommendations. We  improvements in recommendation accuracy by adjusting item recommendations within the proposed framework.

\subsection{LLMs for Recommendation}

There is growing interest in the academic community to harness LLMs for recommendation-related tasks. One research direction explores LLMs within conventional recommendation setup, which typically incorporate user feedback and item metadata~\citep{kang2023llms,hou2023large,yue2023llamarec,dai2023uncovering,bao2023tallrec,harte2023leveraging,sanner2023large}. This includes tasks such as rating prediction~\citep{kang2023llms} and sequential recommendation~\citep{harte2023leveraging,yue2023llamarec,hou2023large}. In such contexts, employing LLMs as recommenders has shown potential, particularly in scenarios with extreme data sparsity~\citep{bao2023tallrec} or during the cold-start phase~\citep{sanner2023large}. However, they often struggle to surpass simpler baseline methods, like non-personalized popularity-based models, in standard recommendation scenarios~\citep{kang2023llms,hou2023large}. Nevertheless, enhancing existing recommender systems with features generated by LLMs has yielded improved performance~\citep{agrawal2023beyond}. Another significant research direction focuses on language-centric recommendation tasks~\citep{he2023large,acharya2023llm,mysore2023large,feng2023large,friedman2023leveraging}. These tasks include generating explanations for recommendations, narrative-based recommendations~\citep{mysore2023large}, and conversational recommendations~\citep{he2023large,feng2023large,friedman2023leveraging}. LLMs exhibit proficient performance in understanding intricate textual inputs, allowing for personalized recommendation outputs. Recent investigations in conversational recommendation demonstrate encouraging outcomes leveraging LLMs, even in zero-shot configurations. Our study employs existing LLMs with minimal additional parameters, implementing the \textit{Reindex-Then-Adapt} framework. Through the reindexing of item content within LLMs and fine-tuning recommendations to align with target data distributions, our framework enhances recommendation accuracy in CRS.

\section{Conclusion}

This study proposes a solution to mitigate distribution misalignments between zero-shot large language models (LLMs) and target recommendation platforms for conversational recommendations. We conceptualize LLMs as Differential Search Index (DSI) models and introduce the \emph{Reindex-Then-Adapt (RTA)} framework. The framework involves converting multi-token item titles into single tokens within LLMs (\texttt{reindex} step) and subsequently adjusting their probability distributions (\texttt{adapt} step). By combining the strengths of LLMs and traditional RecSys, the RTA framework achieves improved recommendation accuracy metrics across various conversational recommendation datasets and adaptation settings.

\bibliographystyle{ACM-Reference-Format}
\bibliography{main}

\newpage
\appendix

\section{More Details of Experiments}

\subsection{Baseline Details}
\label{app:baseline}

We consider four groups of baseline models for comparison. Firstly, we consider some representative traditional \textit{item-based} RecSys models:

\begin{itemize}
    \item \textbf{Popularity}: This method is non-personalized and returns the top-K most popular movies within the related datasets.
    \item \textbf{FISM}~\citep{kabbur2013fism}: A commonly used factored item similarity model for item-based collaborative filtering.
    \item \textbf{SASRec}~\citep{kang2018self}: A competitive self-attention-based sequential recommender system.
\end{itemize}

Secondly, we consider some representative CRS models:

\begin{itemize}
    \item \textbf{ReDIAL}~\citep{li2018redial}: This model is released along with the ReDIAL dataset with an auto-encoder~\citep{sedhain2015autorec}-based recommender.
    \item \textbf{UniCRS}~\citep{wang2022unicrs}: This model uses pre-trained language model, DialoGPT~\citep{zhang2020dialogpt}, with prompt tuning to conduct recommendation and conversation generation tasks respectively. This model is treated as a state-of-the-art CRS models before LLMs~\citep{he2023large}.
\end{itemize}

Thirdly, we consider some dense retrieval models given the connections to document retrieval:

\begin{itemize}
    \item \textbf{SBERT}~\citep{reimers2019sentence}: A modification of the pretrained BERT~\citep{kenton2019bert} network. It uses siamese and triplet network structures to generate semantically meaningful sentence embeddings.
    \item \textbf{Instructor}~\citep{su2022one}: A text embedding model that has been fine-tuned for instructional purposes, which is considered state-of-the-art in dense retrieval tasks.
\end{itemize}

Lastly, we consider some zero-shot open-sourced LLMs as baselines like~\citep{he2023large}, we are using the 7-billion version due to compute burden:

\begin{itemize}
    \item \textbf{MPT-7b}~\citep{MosaicML2023MPT}: A recently released open-sourced LLM released by MosaicML's team trained on 1T tokens.
    \item \textbf{Mistral-7b}~\citep{jiang2023mistral}: A recently released open-sourced Large Language Model with impressive performance on multiple tasks, trained by Mistral AI Team.
    \item \textbf{Llama2-7b}~\citep{touvron2023llama}: A commonly used open-sourced Large Language Model with a wide eco-system support.
\end{itemize}

We also discuss the results from \textbf{GPT-3.5-turbo}~\citep{openai2022chatgpt}, which is a much larger proprietary model that can achieve state-of-the-art CRS performance even in a zero-shot setting~\citep{he2023large}\footnote{In particular, the GPT-3.5-turbo API was called in January 2024 with a temperature setting of 0. This note aims to enhance reproducibility of GPT APIs, considering the continuous updates made by OpenAI over time.}. 

\subsection{Implementation Details} 
\label{app:implementation}
For zero-shot baselines, we configured models based on links from \texttt{huggingface} official model pages for inference on our datasets. Trainable baselines utilized hyperparameters suggested by authors, with a batch size of 256. The learning rate search space is \{1e-3, 1e-4, 1e-5\}, and weight decay is \{0, 1e-6, 1e-4, 1e-2, 1\}. Baselines were trained for 200 epochs, and the best model was selected based on H@10 on the validation set. Reindex and adapt steps of our model followed the same hyper-parameter setup above. For L2R Data and L2I Data, we used the original data mixture without adjusting the sampling ratio. The initial data weights were approximately 98:2, and addressing the data mixture weighting is deferred to future work, as it may enhance recommendation results, though not the primary focus of this paper. For \texttt{Reindex} Step, the RNN we used is a GRU~\citep{chung2014empirical} network, with embedding size as the same as Llama2-7b (i.e.,~4096) and hidden size is 1024. We use the bidirectional single-layer GRU modules. For the \texttt{Adapt} Step, the FISM models are with embedding size 64, and the SASRec models are using embedding size 64, 2 self-attention layers and 2 attention heads. 

\section{Details of Prompts for LLMs}
\label{app:prompts}

\subsection{Prompt(s) for Recommendation}

For LLMs, we follow~\citep{he2023large} to define the recommendation prompts as follows, which can be used to obtain the LLM baseline results and used in our reindex and adapt steps.

For LLM baselines, this prompt is following~\citep{he2023large} with the fuzzy matching method to convert generated recommendation lists into within-dataset item ID lists. In the prompt example, we omit the ``converstaion templates``, which are obtained them from \texttt{FastChat}\footnote{\url{https://github.com/lm-sys/FastChat/blob/1db84d0906196673db361eac50d5aa65180a0ffe/fastchat/conversation.py}} to ensure the zero-shot performance of LLM baselines.

\begin{formal}
  \small
  \textbf{Prompt for LLM Baselines}: \emph{Pretend you are a movie recommender system. 
  I will give you a conversation between a user and you (a recommender system). Based on the conversation, you reply me with 20 movie titles without extra sentences. Here is the conversation:} \textbf{\{\}}
\end{formal}

Here, ``\{\}'' is the placeholder for the conversational context, which is examplified by~\Cref{fig:example}.

For our RTA framework, since the base model is Llama2-7b~\citep{touvron2023llama}, we specify the prompt to make it clear that how we ``reindex'' the generated items. This prompt is exactly ended with ``1. '', for which the original next steps are the tokens for the recommended item titles, such as ``1. Edge of Tomorrow''. However, we record the query embedding ended by ``1. '' and replace the embedding sequence for ``\texttt{Edge of Tomorrow}'' with ``\texttt{|Edge\_of\_Tomorrow|}'' for reindexing. The concrete prompt for the reindex step is:

\begin{formal}
  \small
  \textbf{Prompt for Llama-RTA}: \emph{<s> [INST] Pretend you are a movie recommender system. I will give you a conversation between a user and you (a recommender system). Based on the conversation, you reply me with 20 movie titles without extra sentences. Here is the conversation:} \textbf{\{\}} \emph{[/INST] 1. }
\end{formal}

\subsection{Prompt(s) for Generation}

For the prompt used in case studies~\Cref{fig:example}, we define the prompt as below, where the first placeholder ``\{\}'' is for the conversational context, and the second placeholder ``\{\}'' is for the item recommendation list.

\begin{formal}
  \small
  \textbf{Prompt for Llama-RTA}: \emph{Pretend you are a movie recommender system. I will give you a conversation between a user and you (a recommender system). Based on the conversation, you reply me with 20 movie titles without extra sentences. Here is the conversation:} \textbf{\{\}} \emph{. Please respond the above conversations using the recommended items below, it is better if explaining why they are recommended, but do not list them as bullets. Insert them into your responses:} \textbf{\{\}}
\end{formal}

It is noted that we do not aim to demonstrate the \textit{optimal} generation strategy, but rather provide an example of how the language model framework developed for our recommendation system can also be reused for generative tasks, for the use cases where natural-language responses are required.

\end{document}